# A micrometer-thick oxide film with high thermoelectric performance at temperature ranging from 20-400 K

*Jikun Chen$^{1-3\dagger}$\*, Hongyi Chen$^{2\dagger}$, Feng Hao$^2$, Xinyou Ke$^4$, Nuofu Chen$^5$, Takeaki Yajima$^3$, Yong Jiang$^1$, Xun Shi$^2$, Kexiong Zhou$^2$, Max Döbeli$^6$, Tiansong Zhang$^2$, Binghui Ge$^7$, Hongliang Dong$^8$, Huarong Zeng$^2$, Wenwang Wu$^9$ and Lidong Chen$^{2*}$*

$^1$School of Materials Science and Engineering, University of Science and Technology Beijing, Beijing 100083, China

$^2$Shanghai Institute of Ceramics, Chinese Academy of Sciences, Shanghai 200050, China

$^3$School of Engineering, The University of Tokyo, Tokyo 1138656, Japan

$^4$John A. Paulson School of Engineering and Applied Sciences, Harvard University, Cambridge, Massachusetts 02138, USA

$^5$School of Renewable Energy, North China Electric Power University, Beijing 102206, China

$^6$Ion Beam Physics, ETH Zurich, CH-8093 Zurich, Switzerland

$^7$Beijing National Laboratory for Condensed Matter Physics, Chinese Academy of Sciences, Beijing 100190, China

$^8$Center for High Pressure Science and Technology Advanced Research, Shanghai 201203, China

$^9$Institute of Advanced Structure Technology, Beijing Institute of Technology, Beijing 100081, China

Correspondences: Prof. Lidong Chen (cld@mail.sic.ac.cn) and Prof. Jikun Chen (jikunchen@ustb.edu.cn; jikun@adam.t.u-tokyo.ac.jp). Request for materials: Prof. Jikun Chen (jikunchen@ustb.edu.cn).

*J. Chen is presently at School of Engineering, The University of Tokyo, Tokyo 1138656, Japan.*
*$^\dagger$ J. Chen and H. Chen contribute equally to this work.*

**Thermoelectric (TE) materials achieve localised conversion between thermal and electric energies, and the conversion efficiency is determined by a figure of merit *zT* ($zT=S^2\sigma T\kappa^{-1}$, *S*: Seebeck coefficient, *σ*: electrical conductivity, *T*: absolute temperature, and *κ*: thermal conductivity)[1-16]. Up to date, two-dimensional electron gas (2DEG) related TE materials hold the records for *zT* near room-temperature[1-5]. A sharp increase in *zT* up to ~2.0 was observed previously for superlattice materials such as PbSeTe[1], Bi$_2$Te$_3$/Sb$_2$Te$_3$[2] and**

**$SrNb_{0.2}Ti_{0.8}O_3/SrTiO_3$[3], when the thicknesses of these TE materials were spatially confine within sub-nanometre scale. The two-dimensional confinement of carriers enlarges the density of states near the Fermi energy[3-6] and triggers electron phonon coupling[7-9]. This overcomes the conventional $\sigma$-$S$ trade-off to more independently improve $S$, and thereby further increases thermoelectric power factors ($PF=S^2\sigma$)[6]. Nevertheless, practical applications of the present 2DEG materials for high power energy conversions are impeded by the prerequisite of spatial confinement, as the amount of TE material is insufficient[6,10]. Here, we report similar TE properties to 2DEGs but achieved in $SrNb_{0.2}Ti_{0.8}O_3$ films with thickness within sub-micrometer scale by regulating interfacial and lattice polarizations. High power factor (up to $10^3$ $\mu Wcm^{-1}K^{-2}$) and $zT$ value (up to 1.6) were observed for the film materials near room-temperature and below. Even reckon in the thickness of the substrate, an integrated power factor of both film and substrate approaching to be $10^2$ $\mu Wcm^{-1}K^{-2}$ was achieved in a 2 μm-thick $SrNb_{0.2}Ti_{0.8}O_3$ film grown on a 100 μm-thick $SrTiO_3$ substrate. The dependence of high TE performances on size-confinement is reduced by ~$10^3$ compared to the conventional 2DEG-related TE materials. As-grown oxide films are less toxic and not dependent on large amounts of heavy elements, potentially paving the way towards applications in localised refrigeration and electric power generations.**

Efforts to raise $zT$ value have been focussed on two aspects: reducing the lattice thermal conductivity ($\kappa_{Lattice}$)[10-12] and improving power factors ($PF$)[13]. Since 1990s, $\kappa_{Lattice}$ for many TE material systems has been reduced near the amorphous limit, by using phonon scattering approaches such as fabricating nanostructures[10], enhancing crystal complexities[11] and producing rattling filler-ions within cage-structures[12]. Therefore, it draws more potential to optimise the electronic component: $PF$[13]. Although remarkable $PF$ (~$10^3$μWcm$^{-1}$K$^{-2}$) was observed in SrTiO$_3$-related 2DEGs near room-temperature[3], the preserved large $S$ at a high carrier concentration ($n$) is strongly dependent on two-dimensional (2D) spatial confinement of carriers[1-5]. As observed for SrNb$_{0.2}$Ti$_{0.8}$O$_3$/SrTiO$_3$ superlattice, the Seebeck coefficient is reduced sharply when the thickness of SrNb$_{0.2}$Ti$_{0.8}$O$_3$ exceeds three unit cells (~1.2 nm)[3-5]. Apart from achieving high TE performances, increasing their effective thickness past nanometre scale is also a vital issue to practically applying 2DEG materials for energy conversions.

The polarisations in asymmetric structures provide alternative directions to regulate 2DEGs, not strongly relied on spatial confinement[17-20]. As reported for wurtzite AlGaN/GaN heterostructures, the direction of spontaneous lattice polarisation was ordered by binding with the interfacial polarisation charge through using appropriate epitaxy strategies[17]. The polarisation-induced internal electrostatic field significantly influences the distribution, density and mobility of 2DEGs[17,18,20]. Similarly, polarisation-associated planar charge localizations within LaAlO$_3$/SrTiO$_3$ and La$_{0.5}$Sr$_{0.5}$TiO$_3$/SrTiO$_3$ interfaces were also recognized as an important factor to achieve electronic confinement of 2D carriers to form 2DEGs[7-9]. The observed phonon-drag enhancement in $S$ was reported as a benchmark of electron-phonon coupling (EPC) between the carriers and interfacial polarons within LaAlO$_3$/SrTiO$_3$[7,8].

Following this direction, we herein regulated the TE transportation properties of SrNb$_x$Ti$_{1-x}$O$_3$ by introducing both lattice and interfacial polarisations in a sub-micrometer thick SrNb$_x$Ti$_{1-x}$O$_3$ film coherently grown on a single crystal SrTiO$_3$ (001) substrate. As illustrated by Figure 1a, the unit cell of co-lattice grown SrNb$_{0.2}$Ti$_{0.8}$O$_3$ is compressively distorted by a biaxial in-plane interfacial strain, since it possesses a larger bulk lattice constant ($a_0$=3.96 Å) than SrTiO$_3$ ($a_0$=3.905 Å). This has been demonstrated previously to separate the charge-centre of TiO$_6$ octahedra[22-25] and generate cross-plane Ti$^+$→O$^-$ lattice dipoles (or polarons)[22]. The presence of a coherent interface between electron-doped and intrinsic SrTiO$_3$ is known to result in interfacial polarisation as observed in La$_{0.5}$Sr$_{0.5}$TiO$_3$/SrTiO$_3$ heterostructures[9].

The above-proposed epitaxy has been approached by pulsed laser deposition (PLD), *in situ* monitored by reflection high-energy electron diffraction (RHEED). Figure 1b shows typical RHEED patterns and their intensity oscillations. The same RHEED patterns before and after depositing over $10^3$ unit cells (~440 nm) are observed, indicating no significant changes of the in-plane lattice constant from the surface region. From the high-angle annular dark-field (HAADF) image in transmission electron microscopy as shown in Figure 1c, the co-lattice matched lattice atoms from the film and substrate are seen at their interfacial region, with no detectable diffusion of the Nb element observed (see Figure S4). Further demonstration of the coherent epitaxy is verified by the same in-plane vector of the film and substrate diffraction patterns in reciprocal space mappings (RSM), as one typical example is shown in Figure 1d for SrNb$_{0.2}$Ti$_{0.8}$O$_3$(440 nm)/SrTiO$_3$ (001) and more results are shown in Figure S1 and S2. Figure 1e shows the averaged in-plane and cross-plane lattice displacements ($\varepsilon_{//,\text{Avr.}}$ and $\varepsilon_{\perp,\text{Avr.}}$) derived from their RSM results as shown in Figure S2. The biaxial compressive distortion of the film material is demonstrated by the negative $\varepsilon_{//,\text{Avr.}}$, and the positive $\varepsilon_{\perp,,\text{Avr.}}$ is from the respective cross-plane transverse expansion. With an increasing film thickness and the resultant gradual strain relaxation, both $\varepsilon_{//,\text{Avr.}}$ and $\varepsilon_{\perp,,\text{Avr.}}$ tend towards zero. The interfacial strain completely relaxed for SrNb$_{0.2}$Ti$_{0.8}$O$_3$/LaAlO$_3$ (001) with a large lattice mismatch of ~3.65%.

Figure 2 shows TE performance for SrNb$_{0.2}$Ti$_{0.8}$O$_3$/SrTiO$_3$ (001) and SrNb$_{0.2}$Ti$_{0.8}$O$_3$/LaAlO$_3$ (001) grown under a comparable condition with varied thicknesses ($t_{\text{Film}}$). Linear enhancements are observed for both sheet conductance ($\sigma_{xx}$) and sheet carrier density ($R_H^{-1}$) with increased $t_{\text{Film}}$, as shown by Figure 2a, indicating a constant conductivity, $\sigma = d(\sigma_{xx}) \, dt_{\text{Film}}^{-1}$, and carrier concentration, $n = d(R_H^{-1})dt_{\text{Film}}^{-1}$, associated to the film deposition. Barely annealing the SrTiO$_3$ and LaAlO$_3$ substrate at the deposition atmosphere and temperature (650 °C) for two hours without film depositions does not result in any detectable conductance. A much larger $n$~1.1x10$^{22}$ cm$^{-3}$ associated to deposition is observed for SrNb$_{0.2}$Ti$_{0.8}$O$_3$/SrTiO$_3$ (001), as compared to the ones grown on LaAlO$_3$ (001) substrate under the same condition ($n$~6x10$^{21}$ cm$^{-3}$ contributed by both the Nb dopant and oxygen vacancy of the film material). Similar effect was attributed to both intrinsic and extrinsic reasons as summarized in previous literatures. The intrinsic one includes such as lattice-distortion induced Ti-$t_{2g}$ orbital reconfiguration[21] and formation of 2DEGs by interfacial polarization[7-9] while the extrinsic one is the deposition-associated generation of $V_O^{\cdot\cdot}$ within the SrTiO$_3$ substrate.

Figure 2b shows the directly measured Seebeck coefficient of the SrNb$_{0.2}$Ti$_{0.8}$O$_3$/SrTiO$_3$ (001) and SrNb$_{0.2}$Ti$_{0.8}$O$_3$/LaAlO$_3$ (001) samples ($S_{\text{Film\&Substr.}}$), compared with the previously reported 2DEGs[3-5]. By attributing the enhanced $n$ of SrNb$_{0.2}$Ti$_{0.8}$O$_3$/SrTiO$_3$ (001), compared to SrNb$_{0.2}$Ti$_{0.8}$O$_3$/LaAlO$_3$ (001), completely to the extrinsic $V_O^{\cdot\cdot}$ within the SrTiO$_3$ substrate, we calculated the Seebeck coefficient of the thin film ($S_{\text{Film}}$). More details on derivations are provided in supplementary information (SI): section C. It is worth noting that if there are any intrinsic enhancement in $n$ or the depth distribution of $V_O^{\cdot\cdot}$ is smaller than the thickness of substrate, the practical $S_{\text{Film}}$ will be larger than the calculated ones shown in Figure 2b (See Figure S3b and S3c). In spite of the potential underestimation, significant enhancements in $S_{\text{Film}}$ are found in SrNb$_{0.2}$Ti$_{0.8}$O$_3$/SrTiO$_3$ (001) compared with SrNb$_{0.2}$Ti$_{0.8}$O$_3$/LaAlO$_3$ (001) and the reported bulk SrNb$_{0.2}$Ti$_{0.8}$O$_3$[26]. Increasing $t_{\text{Film}}$ from 45 nm to 2.2 μm reduces $S_{\text{Film}}$ by ~200 μVK$^{-1}$ while the 2.2 μm thick film maintains $S_{\text{Film}}$ approaching to be ~170 μVK$^{-1}$. This is in contrast to a much sharper decrease in $S$ by ~400 μVK$^{-1}$ reported in SrNb$_{0.2}$Ti$_{0.8}$O$_3$ 2DEGs[3-5] when the $t_{\text{Film}}$ only increased from one to three unit cells (with similar $S_{\text{Film}}$ to the bulk value at $t_{\text{Film}}$~1.2 nm). Further consistency was achieved from the localised $S$ near-surface regions with response depth as ~100 nm of an 800 nm-thick films measured by using nanometre-scaled heating source in atomic force microscopy ($S_{\text{Near-Surf.}}$= 290 ± 18

μVK$^{-1}$, as shown by Figure 2c). More details for the localized characterization of *S* are given in SI: Section E and ref. S5.

The resultant power factors for the films are from $10^2$ to $10^3$ μWcm$^{-1}$K$^{-2}$ while their *zT* are estimated as 0.3 to 1.6 using similar estimations according to ref 3 ($\kappa = \kappa_{Lattice} + \kappa_{Carrier}$, where $\kappa_{Lattice}$ ~12 Wm$^{-1}$k$^{-1}$, and $\kappa_{Carrier} = L_0 \sigma T$, $L_0 = 2.45 \times 10^{-8}$ WΩ$^{-1}$K$^{-2}$) as shown in Figure 2d. These TE performances are 2 orders larger than the SrNb$_{0.2}$Ti$_{0.8}$O$_3$ bulk material, and have been verified by extensive experimental labours as more detailed results provided in SI: section D-F. Vice versa, performing thermoshock or annealing in vacuum to eliminate the interfacial coherency and relax the interfacial strain significantly reduces both *S* and *σ* as shown by Figure 3a and S7.

To further saturate the generation of $V_O^{\cdot\cdot}$ within the substrate and reduce its influence to the directly measured TE performances, we deposited a 2 μm thick SrNb$_{0.2}$Ti$_{0.8}$O$_3$ film grown on 100 μm thick SrTiO$_3$ (001) substrate under the same condition. As shown in Figure 3b, the $\sigma_{xx}$ measured for SrNb$_{0.2}$Ti$_{0.8}$O$_3$ ($t_{Film}$=2 μm)/SrTiO$_3$ ($t_{Substr.}$=100 μm) overlaps with the derived film contribution ($\sigma_{xx,Film}$, and its derivation is shown in SI: Section C) of SrNb$_{0.2}$Ti$_{0.8}$O$_3$ ($t_{Film}$=2.2 μm)/SrTiO$_3$ ($t_{Substr.}$=1 mm) at 100 to 300 K. It reveals a significantly reduced proportion of $\sigma_{xx}$ from the substrate associated to the saturation of $V_O^{\cdot\cdot}$ concentration at the reduced amount of substrate material. This provides further opportunities to more directly characterize the TE properties associated to the film material. Figure 3c shows the temperature dependant *S* and electrical conductivity (*σ*) of the SrNb$_{0.2}$Ti$_{0.8}$O$_3$ (2 μm) / SrTiO$_3$ (100 μm) sample measured as a bulk ($t = t_{Film} + t_{Subst.}$ in calculation of *σ*). Similar to the reported SrNb$_x$Ti$_{1-x}$O$_3$ single crystalline materials[27,29,30], elevating the temperature enhances $S$ and decreases $\sigma$. Even reckon in a 50 times thicker substrate, a bulk *PF* (integrating both film and substrate) approaching to be ~$10^2$ µWcm$^{-1}$K$^{-2}$ is achievable at low temperatures (see Figure 3d).

In contrast to the previous 2DEG-related SrNb$_{0.2}$Ti$_{0.8}$O$_3$[3-5], the enhanced Seebeck coefficient for SrNb$_{0.2}$Ti$_{0.8}$O$_3$/SrTiO$_3$ (001) in this work is not strongly related to spatial confinements. The present achieved performance is comparable with the one for oxygen annealed LaAlO$_3$/SrTiO$_3$, in which case similar magnitudes of $S$ (~600 µVK$^{-1}$) at large $n_{2D}$ (the order of high $10^{13}$ up to $10^{14}$ cm$^{-2}$) was achieved by the interfacial polarization-induced electron-phonon coupling[7,8]. The high relative permittivity ($\varepsilon_r$: ~$10^3$) and ferroelectric nature of strain-distorted SrTiO$_3$[23] makes an electronic 2D confinement of carriers to be practicable, when the cross-plane lattice polarons reaches an orderly alignment (see more discussions in SI: Section G). The aligned cross-plane lattice polarizations were observed previously for AlGaN/GaN heterostructures[17,18,20], and are thermodynamically favoured by binding with the accumulated positive charge at the interface under appropriate kinetic processes.

This understanding is in agreement with the larger Seebeck coefficient observed for the thinner SrNb$_{0.2}$Ti$_{0.8}$O$_3$/SrTiO$_3$ (001) with better preserved lattice distortions that produces the lattice polarization. In addition, improving the interfacial polarization by producing a thin buffering layer of BaTiO$_3$ with a larger relative permittivity was observed to further enhance $S_{Film\&Substr.}$ for ~10%, while maintaining a similar film conductance. This was observed in SrNb$_{0.2}$Ti$_{0.8}$O$_3$(400 nm)/BaTiO$_3$(20

nm)/SrTiO$_3$ (001): $S= -540$ μVK$^{-1}$, $\sigma= 4.3\times10^5$ Sm$^{-1}$ at room temperature. Vice versa, the compressive strain alone does not result in similar high TE performance as confirmed in compressive-strained SrNb$_x$Ti$_{1-x}$O$_3$ grown on DyScO$_3$ (001) or KTaO$_3$ (001) substrates as shown in Figure S9. These results reveal a potential coupling between a lattice and an interfacial polarisation for reaching an orderly alignment, which achieves enhancement in TE performance similarly to the reported 2DEG-SrNb$_x$Ti$_{1-x}$O$_3$[3-5] by electronic 2D confinement rather than size-confinement.

In summary, similar TE performance to 2DEG-related SrNb$_x$Ti$_{1-x}$O$_3$ has been achieved in a sub-micrometer scale within the room-temperature and low-temperature ranges. Such high performance is related to both the distortion-induced lattice polarisation and the interfacial polarisation, while the dependence on the size-confinement is significantly reduced. The minimal usages of heavy elements and low toxicity can pave the way towards practical applications for energy harvesting and localized cooling.

**Acknowledgments**

This work is supported by the Fundamental Research Funds for the Central Universities (USTB), National Natural Science Foundation of China (No. 51602022, No. 61674013 and No. 11374332), the key research program of Chinese Academy of Sciences (Grant No. KGZD-EW-T06), and the research grant from Shanghai government (No. 15JC1400301). JC also appreciate Japanese Society for the Promotion of Science (Fellowship ID: P15363). We appreciate helpful discussions with Prof. Akira Toriumi from The University of Tokyo (Japan), Prof. Jian Shi from Rensselaer Polytechnic Institute (USA), Prof. Rafael Jaramillo from Massachusetts Institute of Technology (USA), Prof. Xuchun Gui from Sun Yat-sen University (China), Prof. Renkui Zheng from Shanghai Institute of Ceramics, Chinese Academy of Sciences (China). We also acknowledge Mr. Charles Plumridge from The University of Tokyo for revising this manuscript.

**Competing Interests**

We declare that we do not have any competing financial interest.

**Additional Information:** Supplementary Information is available for this manuscript.

**Correspondences:** Correspondence should be addressed to: Prof. Lidong Chen (cld@mail.sic.ac.cn) and Prof. Jikun Chen (jikunchen@ustb.edu.cn). Request for materials should contact Prof. Jikun Chen (jikunchen@ustb.edu.cn).

**Contributions**

JC proposed the idea, planed for the experiments, developed the film deposition strategy, performed film growth, and wrote the manuscript assisted by LC; HC and FH performed the verification experiments, advised by LC, JC, XS and HZ; HC, FH, KZ and TZ characterized the transportation performances; JC, XK and WW contributed to the RSM measurement and analysis; BG, FH and HD contributed to the TEM measurement; JC and NC proposed the understanding shown in Figure S9 assisted by LC and XK; JC, LC, NC, YJ, YT and MD provided experimental support and useful discussions.

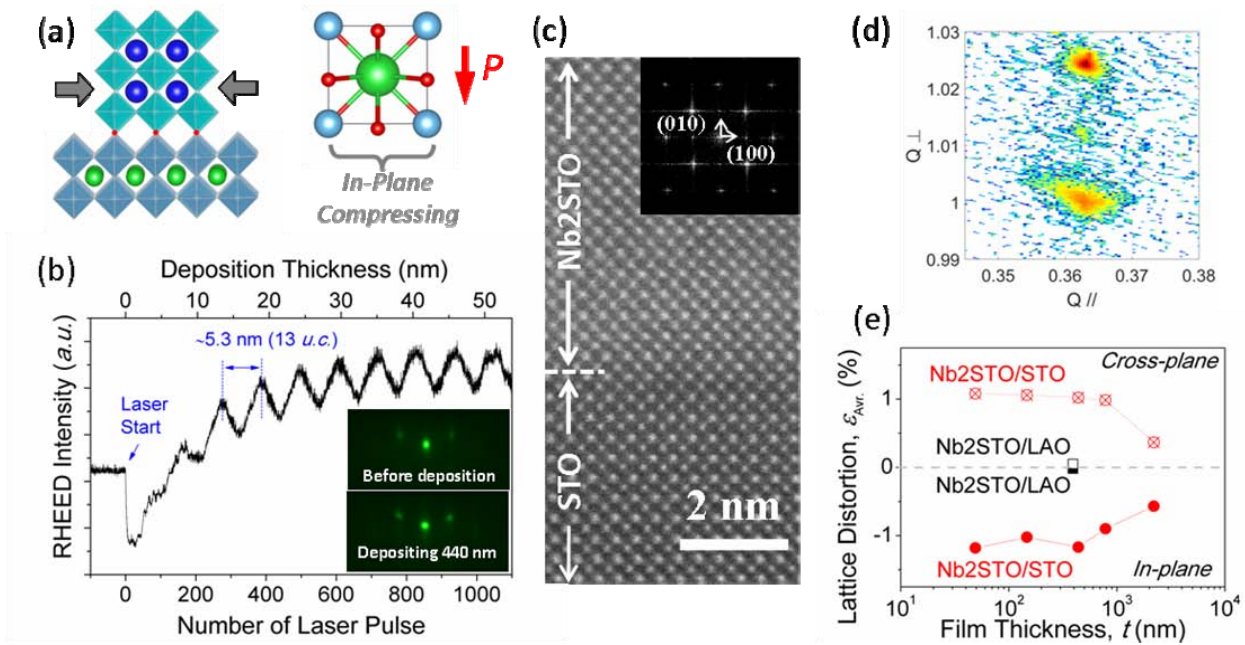

**Figure 1.** **(a)** Schematic illustration of the co-lattice epitaxy of SrNb$_x$Ti$_{1-x}$O$_3$ on single crystal SrTiO$_3$ (001) substrate and the cross-plan polarization induced by in-plane compressive distortion. **(b)** Reflection high-energy electron diffraction (RHEED)

patterns and intensity oscillations for the present deposition of SrNb$_x$Ti$_{1-x}$O$_3$ on a SrTiO$_3$ (001) substrate **(c)** High-angle annular dark-field (HAADF) image of the interface for the SrNb$_{0.2}$Ti$_{0.8}$O$_3$/SrTiO$_3$ (001). **(d)** Reciprocal space mapping (RSM) of SrNb$_{0.2}$Ti$_{0.8}$O$_3$/SrTiO$_3$ (001) samples with a film thickness of 440 nm, where $Q_{//}$ and $Q_\perp$ represent the in-plane and cross-plane reciprocal space vector, respectively. The diffraction patterns for the film, $I_{Film(Q_{//},Q_\perp)}$, and substrate, $I_{Film(Q_{//},Q_\perp)}$, are located at the same in-plane reciprocal space vector $Q_{//}$, indicating their in-plane lattice constant is similar. In contrast, a smaller cross-plane reciprocal space vector $Q_\perp$ is observed for the film compared with the substrate, owing to a larger cross-plane lattice constant. **(e)** The averaged in-plane and cross-plane lattice displacements ($\varepsilon_{Avr.}$) for SrNb$_{0.2}$Ti$_{0.8}$O$_3$/SrTiO$_3$ (001) samples with different film thicknesses, estimated from their RSM as shown in Figure 1d and Figure S2.

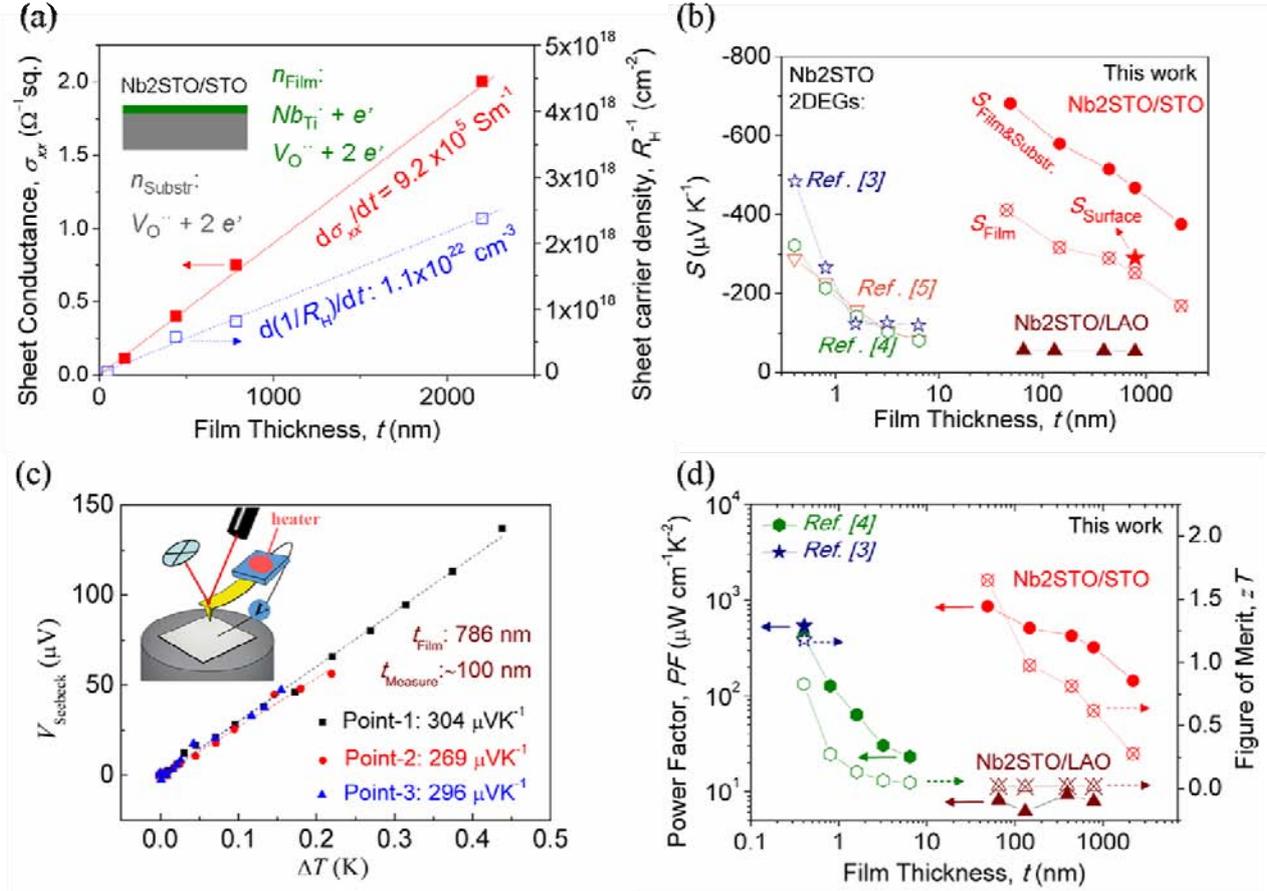

**Figure 2.** **(a)** Sheet conductance ($\sigma_{xx}$) and Sheet carrier density (the reciprocal of Hall resistance, $R_H^{-1}$) as a function of film thickness ($t$) at room-temperature for SrNb$_{0.2}$Ti$_{0.8}$O$_3$/SrTiO$_3$ (001) and SrNb$_{0.2}$Ti$_{0.8}$O$_3$/LaAlO$_3$ (001). **(b)** Directly measured Seebeck coefficient for film and substrate ($S_{Film\&Substr.}$), the derived one for film materials ($S_{Film}$) and the localized $S$ measured within 100 nm depth from the near-surface region of the film ($S_{Surface}$) using nanometre-scale heating source in atomic force microscopy. **(c)** Seebeck voltage ($V_{Seebeck}$) vs. temperature difference ($\Delta T$) at the near-surface region of the film during the localized characterizations of $S$. **(d)** The estimated power factor and figure of merit. The reported $S$, $PF$ and $zT$ for 2DEG-SrNb$_{0.2}$Ti$_{0.8}$O$_3$[3-5] are plotted to compare. The detailed approach to separate the film and substrate contribution is provided in SI: Section C.

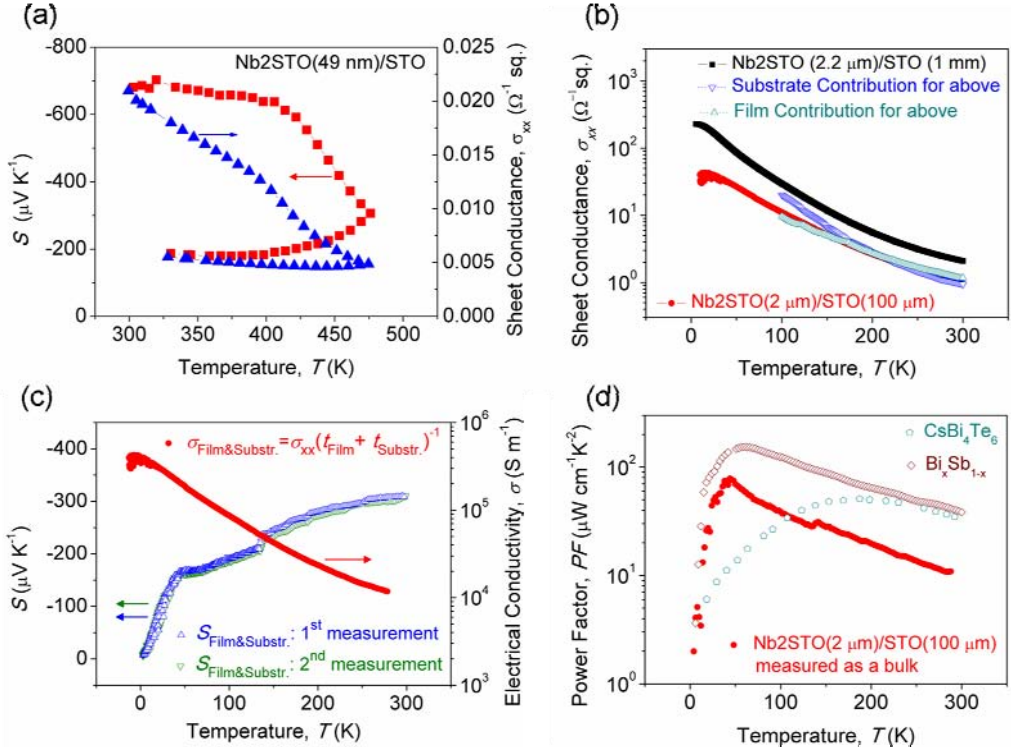

**Figure 3.** **(a)** The Seebeck coefficient and sheet conductance of a 49 nm thick SrNb$_{0.2}$Ti$_{0.8}$O$_3$/SrTiO$_3$ (001) measured with raising temperature up to 200 °C followed by cooling down technique with a constant speed of 0.3 °C min$^{-1}$ in vacuum. **(b)** Sheet conductance of a 2.2 μm thick SrNb$_{0.2}$Ti$_{0.8}$O$_3$ film grown on 1 mm thick SrTiO$_3$ (001) substrate and a 2 μm thick SrNb$_{0.2}$Ti$_{0.8}$O$_3$ film grown on 100 μm thick SrTiO$_3$ (001) substrate as a function of temperature. The contribution from both film and substrate to the sheet conductance is separated for the SrNb$_{0.2}$Ti$_{0.8}$O$_3$ (2.2 μm)/SrTiO$_3$ (1 mm) sample (more details are provided in SI: Section C). The overlapping between film contributed sheet conductance (dark cyan triangle) and that measured for SrNb$_{0.2}$Ti$_{0.8}$O$_3$ (2 μm)/SrTiO$_3$ (100 μm) sample (red circular) indicates a small conductance contributed by the substrate of the latter sample. **(c)** The Seebeck coefficient, electrical conductivity and **(d)** power factor of the SrNb$_{0.2}$Ti$_{0.8}$O$_3$ (2 μm)/SrTiO$_3$ (100 μm) sample measured as bulk material (calculation reckon in the thickness of the substrate), compared to conventional thermoelectric materials for low-temperature applications, such as $Bi_xSb_{1-x}$[15] and $CsBi_4Te_6$[16].

# Supporting Information

# A micrometer-thick oxide film with high thermoelectric performance at temperature ranging from 20-400 K

*Jikun Chen[1-3]\*, Hongyi Chen[2], Feng Hao[2], Xinyou Ke[4], Nuofu Chen[5], Takeaki Yajima[3], Yong Jiang[1], Xun Shi[2], Kexiong Zhou[2], Max Döbeli[6], Tiansong Zhang[2], Binghui Ge[7], Hongliang Dong[8], Huarong Zeng[2], Wenwang Wu[9] and Lidong Chen[2]\**

[1] School of Materials Science and Engineering, University of Science and Technology Beijing, Beijing 100083, China

[2] Shanghai Institute of Ceramics, Chinese Academy of Sciences, Shanghai 200050, China

[3] School of Engineering, The University of Tokyo, Tokyo 1138656, Japan

[4] John A. Paulson School of Engineering and Applied Sciences, Harvard University, Cambridge, Massachusetts 02138, USA

[5] School of Renewable Energy, North China Electric Power University, Beijing 102206, China

[6] Ion Beam Physics, ETH Zurich, CH-8093 Zurich, Switzerland

[7] Beijing National Laboratory for Condensed Matter Physics, Chinese Academy of Sciences, Beijing 100190, China

[8] Center for High Pressure Science and Technology Advanced Research, Shanghai 201203, China

[9] Institute of Advanced Structure Technology, Beijing Institute of Technology, Beijing 100081, China

Correspondences: Prof. Lidong Chen (cld@mail.sic.ac.cn) and Prof. Jikun Chen (jikunchen@ustb.edu.cn).

Request for materials: Prof. Jikun Chen (jikunchen@ustb.edu.cn).

**Section A: Methods**

Thin films were grown on SrTiO$_3$ (001) and LaAlO$_3$ (001) single crystal substrates by pulsed laser deposition (PLD) by using ceramic targets with nominal compositions of SrNb$_{0.2}$Ti$_{0.8}$O$_3$. The laser ablation and background conditions were kept the same for all depositions, while the substrates were heated to 650 °C during the deposition, similar to ref S1. High-angle annular dark-field (HAADF) and annular bright-field (ABF) scanning transmission electron m microscopy (STEM) experimental techniques were carried out on

JEM-ARM 200F TEM operated at 200 kV with a cold field emission gun and aberration correctors for both probe-forming and imaging lenses. The crystal structures were characterized by X-ray diffraction (XRD) and reciprocal space mapping (RSM). The diffraction patterns of [114] reciprocal space vectors from the film and substrate were projected at [110] and [001], representing the in-plane and cross-plane reciprocal space vector ($Q_{//}$ and $Q_{\perp}$), respectively. The RSM result from [114] diffraction pattern demonstrates the in-plane lock between film and substrate in the other in-plane direction of [100]. More results for XRD and RSM are further shown in Figure S1 and S2.

The electrical conductivities and Seebeck coefficients of as-grown thin films within temperature ranging from 300 to 500 K were measured by a self-developed system according to ref. S2. Measurement of standard samples from the present setup and the commercialized Ulvac ZEM-3 system shows consistent results. In addition, room-temperature performance by using the present setup for thin film samples is in agreement with those measured by Physical Property Measurement System (PPSM). The $S$ and $\sigma$ for $SrNb_{0.2}Ti_{0.6}O_3/LaAlO_3$ (001) are similar to the previous reported $SrNb_{0.2}Ti_{0.6}O_3$ thin films (ref. S3 and ref. 27). The low-temperature transportation properties from 5 to 300 K were characterized by PPMS (Quantum Design) under a high vacuum. The Hall resistance ($R_H$) and electrical conductivity ($\sigma$) were measured by using the Hall and resistivity options for an AC transport on Quantum Design PPMS.

The localised measurement of Seebeck coefficient at the near-surface region of the film was performed according to ref. S5 on the commercial atomic force microscope (SPA 400, SPI3800N, Seiko Inc. Japan). In brief, the miniature heating parts is composed of copper bar wrapped with the constantan wire were set on the backside of the conductive cantilever. Under a direct current (DC) voltage, Joule heat is generated on the constantan wire with constant resistivity at room-temperature and further transports to the conductive AFM tip along the cantilever. The heated tip results in a local temperature rise at the nanoscale contact region when it is brought in contact with the thermoelectric sample surface. As a result, a small temperature difference ($\Delta T$) appears between the heated nano-contact region and non-heated region of the sample surface, and gives rise to a localised Seebeck voltage ($V_{\text{Seebeck}}$). The localised Seebeck coefficient ($S_{\text{Surface}}$) is obtained by $S_{\text{Surface}}=V_{\text{Seebeck}}/\Delta T$, where $\Delta T$ is calibrated by reference samples of $CoSb_3$. The used conductive probe was a platinum-/titanium-coated silicon cantilever with spring constant of 4.5 N/m and a resonance frequency of 70 kHz. The DC voltage was a regulated DC power supply (Model YJ56), and the detection of DC Seebeck was by a digit precision multimeter (Tektronix DMM4050).

**Section B: XRD and RSM results**

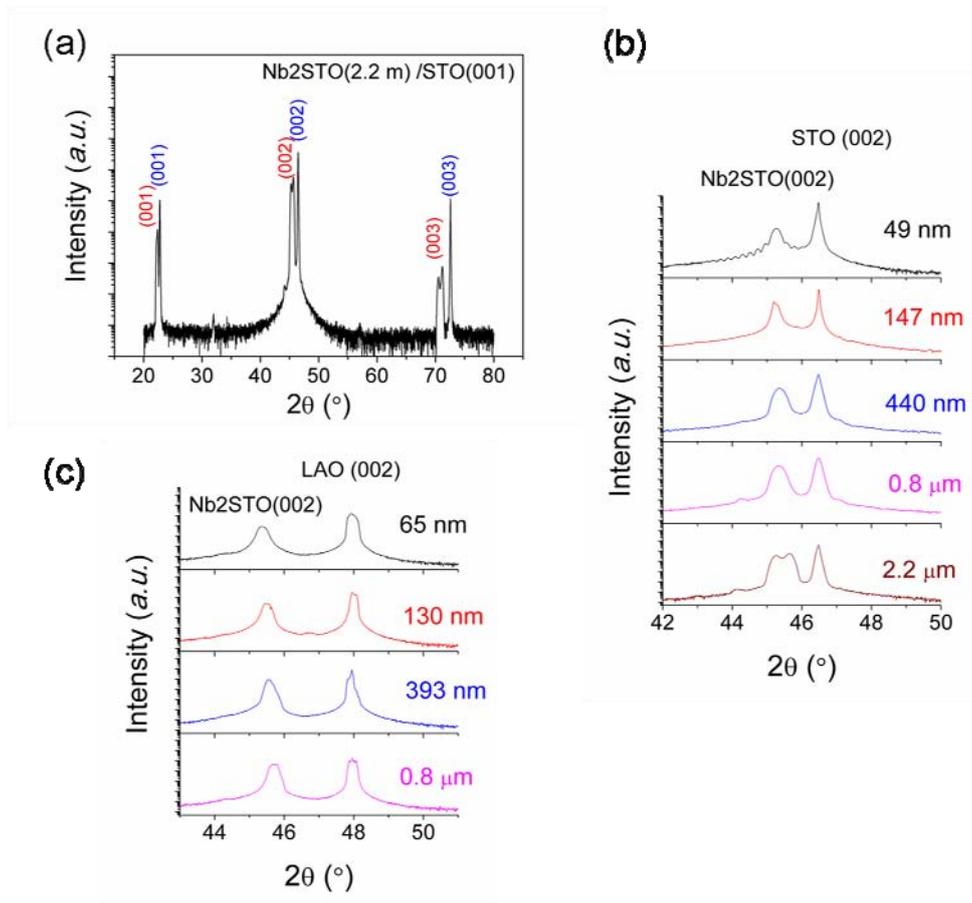

**Figure S1.** **(a)** Representative examples of the XRD patterns (θ-2θ scan) for **(a)** SrNb$_{0.2}$Ti$_{0.8}$O$_3$/SrTiO$_3$ (001) with a film thickness of 2.2 μm, **(b)** SrNb$_{0.2}$Ti$_{0.8}$O$_3$/SrTiO$_3$ (001) with different film thicknesses, **(c)** SrNb$_{0.2}$Ti$_{0.8}$O$_3$/LaAlO$_3$ (001) with different film thicknesses. The film and substrate show the same crystal structure and orientation, since the diffraction peaks for the film are present beside those for the substrate (**a**). A gradual right-shift and broadening of film diffraction peaks with an increased deposition thickness was observed in (**b**), which indicate a gradually varied cross-plane lattice constant with the relaxation of interfacial strains.

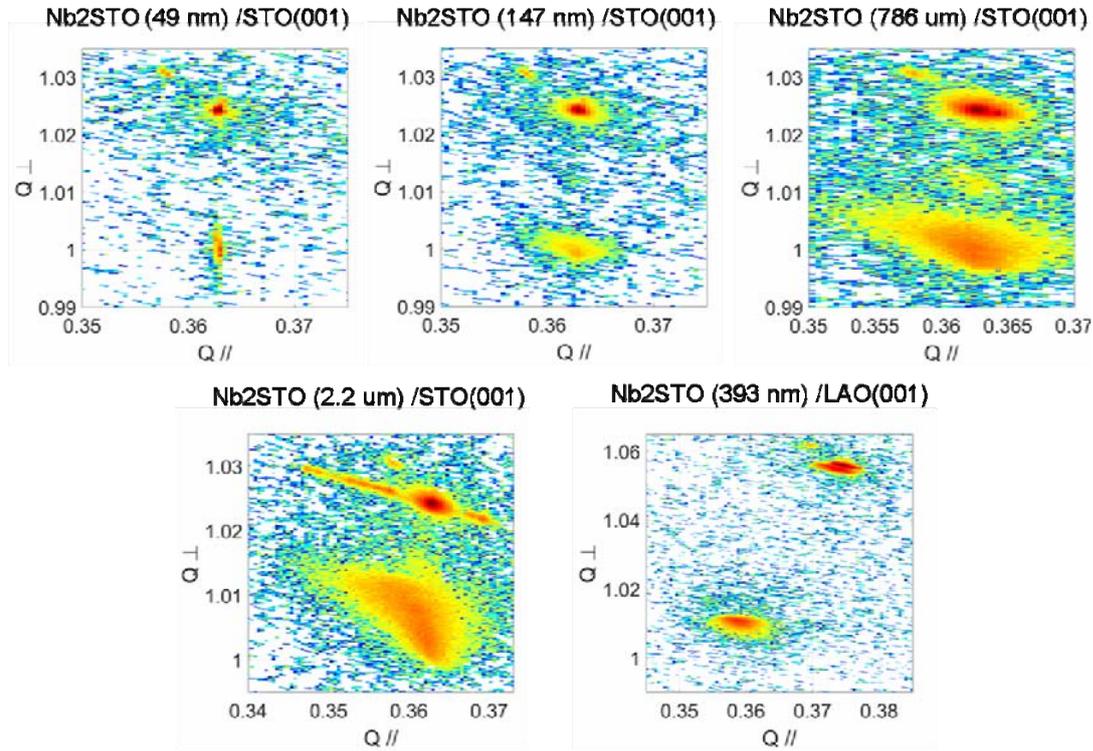

**Figure S2.** RSM results for *as-grown* SrNb$_{0.2}$Ti$_{0.8}$O$_3$/SrTiO$_3$ (001) and SrNb$_{0.2}$Ti$_{0.8}$O$_3$/LaAlO$_3$ (001) with different deposition thicknesses. In each sub-figure, the upper and lower diffraction patterns are the reciprocal space vectors of [114] from the substrate and film, respectively. Thin film grown on SrTiO$_3$ (001) substrate shows a similar projection of diffraction patterns in $Q_{//}$ compared with the substrate, indicating their similar in-plane lattice constants from coherent epitaxy. A smaller $Q_\perp$ observed for the film indicates its larger cross-plane lattice constant than the substrate. In contrast, the SrNb$_{0.2}$Ti$_{0.8}$O$_3$/LaAlO$_3$ (001) shows smaller $Q_{//}$ and $Q_\perp$ for the film compared with the substrate, indicating larger lattice constants for both in-plane and cross-plane directions. It reveals that the interfacial strain is not preserved and this is from a large lattice mismatch (3.65%). The averaged in-plane and cross-plane lattice displacements ($\varepsilon_{//,\text{Avr.}}$ and $\varepsilon_{\perp,\text{Avr.}}$) derived from the film pattern, $I_{Film(Q_{//},Q_\perp)}$, in the RSM, by:

$$\varepsilon_{\|(\perp),Avr.} = \frac{\int_{Q_{//(\perp)}} [(\frac{1/Q_{//(\perp)} - 1/Q_{//0(\perp 0)}}{1/Q_{//(\perp)}}) \times \int_{Q_{\perp(//)}} I_{Film}(Q_{//},Q_{\perp})dQ_{\perp(//)}]dQ_{//(\perp)}}{\iint_{Q_{//},Q_{\perp}} I_{Film}(Q_{//},Q_{\perp})dQ_{//}dQ_{\perp}}$$

**Section C: Separating the transportation properties contributed by the film and the substrate.**

*Room Temperature:* Neglecting the contact resistance between the film and substrate, the practical measured transportation performance is considered as the parallel between film ($SrNb_{0.2}Ti_{0.8}O_3$) and substrate (reduced $SrTiO_3$). From the deposition associated carrier concentration ($n=dR_H^{-1}dt^{-1}$: ~1.1x10$^{22}$ cm$^{-3}$), the total number of carrier when depositing film with thickness of $t_{Film}$

$$N_{xx}=nt_{Film}$$

Similarly, from the deposition associated electrical conductivity ($\sigma=d\sigma_{xx}dt^{-1}$: ~9.2 x 10$^5$ Sm$^{-1}$), the total sheet conductance when depositing film with thickness of $t_{Film}$

$$\sigma_{xx}=\sigma t_{Film}$$

Assuming the proportion of carrier allocated to film is $x$, and thereby the one allocated to substrate is 1-$x$. For the estimation in the main part of the manuscript, we assume $x=n_{Nb2STO/LAO} - n_{Nb2STO/STO}$. The carrier concentration of film material is $n_{Film}=xn$, while for substrate material is $n_{Substr.}=(1-x)nt_{Film}t_{Substr}^{-1}$. The substrate is considered as a conventional electron doped SrTiO$_3$ and its Seebeck coefficient ($S_{substr.}$) is related to $n_{Substr.}$ by $S_{substr.}$ =$A\log(n_{Substr.})+B$ summarized in ref. 3 and ref. 27 (see the plot shown by Figure S3a).

The sheet conductance of substrate is $\sigma_{xx,Substrate}=(1-x)nt_{Film}t_{Substr.}^{-1}e\mu_{Substr.}$; where $\mu_{Substr.}$ is around 6 cm$^2$V$^{-1}$S$^{-1}$ for a lightly doped SrTiO$_3$ similar to the treatment report by ref. 3. Therefore, the sheet conductance of film is calculated by $\sigma_{xx,Film}=\sigma t_{Film}-\sigma_{xx,Substr.}e\mu_{Substr.}$, while the conductivity of the film is further calculated by: $\sigma_{,Film}=\sigma_{xx,Film} t_{Film}$. The The mobility of the thin film is calculated from the relationship of $\sigma_{xx,Film}=n_{Film}e\mu_{Film}$. The mobility of the film can be further calculated by $\mu_{Film}=\sigma_{xx,Film}(en_{Film})^{-1}$. The calculated $\mu_{Film}$ for SrNb$_{0.2}$Ti$_{0.8}$O$_3$/SrTiO$_3$ (001) at room temperature is 5.0 cm$^2$V$^{-1}$S$^{-1}$. Since $S_{Film\&Substr.}=(S_{Film}\sigma_{xx,Film}+S_{Substrate}\sigma_{xx,Substrate})\sigma_{xx,Film\&Substr.}^{-1}$, the Seebeck coefficient of film can be calculated by $S_{Film}=[S_{Measure}\sigma t_{Film}-S_{substrate}\sigma_{xx,Substrate}(\sigma t_{Film})^{-1}][\sigma_{xx,Film}(\sigma t_{Film})^{-1}]^{-1}$.

To avoid overestimation of the TE performance of film material in the main part of the manuscript, we attributed the enhanced carrier density observed in SrNb$_{0.2}$Ti$_{0.8}$O$_3$/SrTiO$_3$ (001) as compared to SrNb$_{0.2}$Ti$_{0.8}$O$_3$/LaAlO$_3$ (001) completely to the extrinsic generation of oxygen vacancy from the substrate. Nevertheless, it is worth noting that this calculation will underestimate the practical magnitude of $S_{Film}$ when there is a proportion of intrinsic enhancement of carrier density by such as strain-induced electron configuration that triggers metal to insulator transitions. This is more clearly shown by Figure S3b. Increasing the proportion of carriers allocated to the film ($x$) results in larger magnitudes of the calculated $S_{Film}$. In addition, the calculated $S_{Film}$ as shown in Figure 2c will be underestimated when the oxygen vacancy distributed shallower than the depth of the substrate thickness (see Figure S3c).

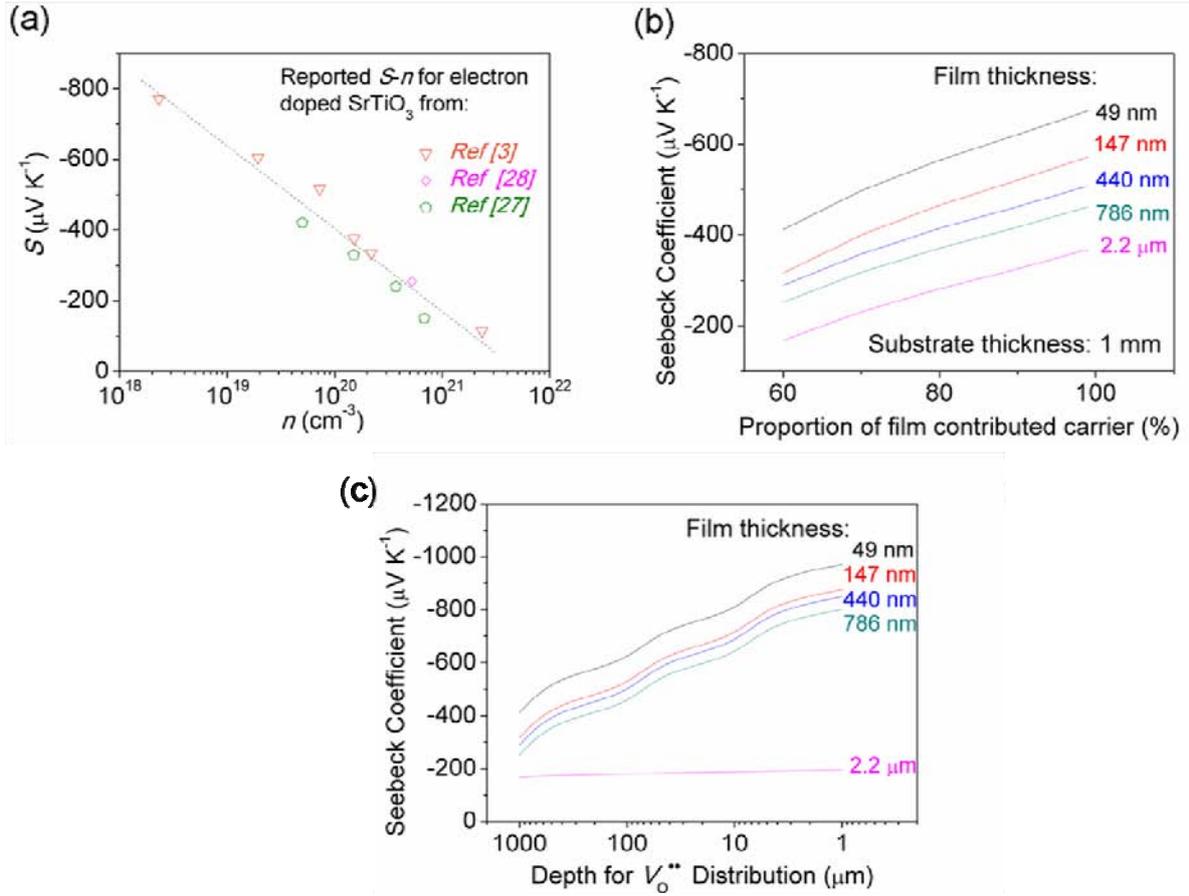

**Figure S3. (a)** Relationship between the Seebeck coefficient ($S$) and the carrier concentration ($n$). **(b) (c)** Derived Seebeck coefficient of the $SrNb_{0.2}Ti_{0.8}O_3$ film material (grown on 1 mm thick $SrTi_0O_3$ substrate) as a function of **(b)** the proportion of carriers allocated to the film material and **(c)** the depth for distribution of the oxygen vacancy within the substrate. Figure S5a shows that the calculated $S_{Film}$ in Figure 2c will be underestimated when there is a proportion of intrinsic enhancement of carrier density by such as strain-induced electron configuration that triggers metal to insulator transitions. Figure S3b shows that the calculated $S_{Film}$ in Figure 2c will be underestimated when the oxygen vacancy distributed shallower than the depth of the substrate thickness.

***Temperature dependence:*** The lightly doped single crystalline $SrTiO_3$ under the same carrier concentration possesses similar temperature dependent transportation behaviors[29,30]. We uses the $\sigma_{xx,Substr.}$ at room temperature times the temperature dependence reported for single crystalline electron doped SrTiO$_3$ with similar carrier concentration ($\sim 10^{19}$ cm$^{-3}$)[30] to estimate its temperature dependence for SrNb$_{0.2}$Ti$_{0.8}$O$_3$ (2.2 μm)/SrTiO$_3$ (1 mm) as shown in Figure 3a. For the situation of for SrNb$_{0.2}$Ti$_{0.8}$O$_3$ (2 μm)/SrTiO$_3$ (100 μm), a significant reduction in sheet resistance is observed as compared with SrNb$_{0.2}$Ti$_{0.8}$O$_3$ (2.2 μm)/SrTiO$_3$ (1 mm), which derivates from the linear $\sigma_{xx}$-$t_{Film}$ relationship as shown in Figure 2a. It is worth noting that the practically measured $\sigma_{xx}$ for SrNb$_{0.2}$Ti$_{0.8}$O$_3$ (2 μm)/SrTiO$_3$ (100 μm) overlaps with the film contributed $\sigma_{xx,Film}$ separated for SrNb$_{0.2}$Ti$_{0.8}$O$_3$ (2.2 μm)/SrTiO$_3$ (1 mm). These results indicate that the oxygen vacancy induced electron doping of the substrate saturated at $\sim 10^{19}$ cm$^{-3}$, and the influence of the substrate to the practically measured sheet conductance and Seebeck coefficient is small at temperature higher than 100 K.

**Section D: Excluding potential misinterpretations for the characterization of *S* and *σ***

To exclude the effect from the surface layer of the film during the measurement of *S* and *σ*, we referred to the model proposed by Danielson[S4] and circled the four contacting electrodes together with the film and substrate using a conductive paste to reproduce the measurement (see Figure S4a). The results are similar to those measured by using the standard way: variation in *σ* (<5%, see Figure S4b) and variation in *S* (<2%, see Figure S4c). This demonstrates that the observed *S* and *σ* are not associated to the surface layer of the thin film.

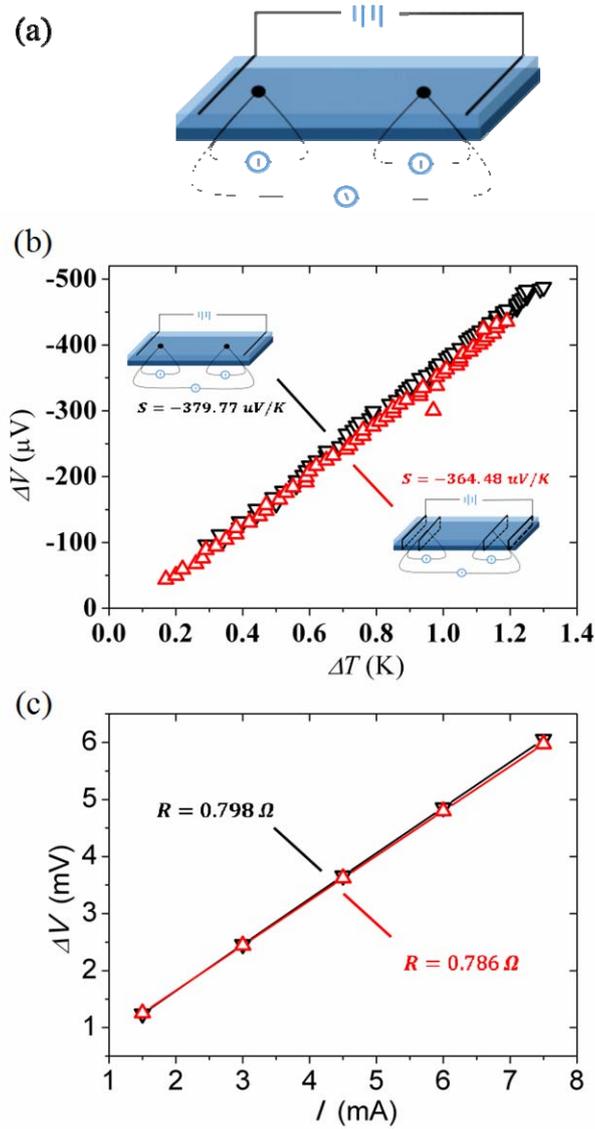

**Figure S4. (a)** Schematic illustration for the present measurement of $S$ and $\sigma$. **(b)**, **(c)** Representative $\varDelta V$-$\varDelta T$, $\varDelta V$-$\varDelta I$ characterizations for measuring $S$ and $\sigma$ using standard approach (black line) and after circling the four contact electrodes together with the film and substrate using conductive paste (red line) based on the model proposed by Danielson[S4].

The potential diffusion from film to substrate was investigated. As shown in Figure S5, no detectable diffusion of Nb was observed at the interfacial region between $SrNb_xTi_{1-x}O_3$ and $SrTiO_3$. It confirms that the presently observed high TE performance for $SrNb_xTi_{1-x}O_3/SrTiO_3$

(001) is not associated to interfacial diffusions of the Nb element.

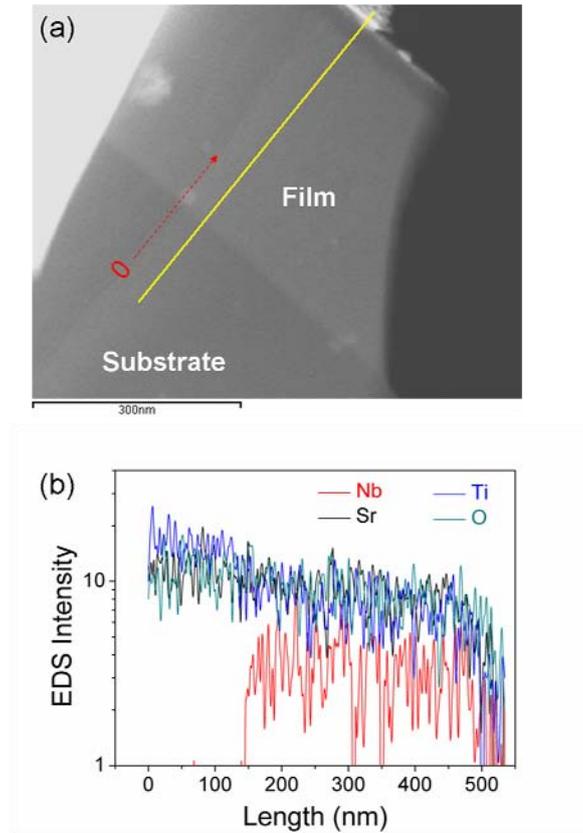

**Figure S5.** Transmission electron microscope (TEM) image and energy-dispersive X-ray spectroscopy (EDS) line scan of the elements across the interface between *as-grown* SrNb$_x$Ti$_{1-x}$O$_3$ film and SrTiO$_3$ substrate. No detectable diffusion of Nb was observed.

**Section E: Localized Seebeck coefficient measured from the near-surface region of the sub-micrometer films using nano-scale heating source in atomic force microscopy (AFM)**

We also performed localized characterization of *S* from the near-surface region (responding depth is ~100 nm) from a 0.8 μm thick SrNb$_{0.2}$Ti$_{0.8}$O$_3$/SrTiO$_3$ (001) by using nano-scaled heating source in atomic force microscopy (AFM) as reported by ref S5. As shown in Figure S6a, a heated probe cantilever was contact with the surface of the film within a localized region of ~$10^1$ nm$^2$ and a temperature difference within 100 nm was established. The temperature gradient was varied by applying different magnitude of heating voltage ($V_{heat}$), and the voltage from Seebeck effect ($V_{seebeck}$) at each $V_{heat}$ was measured. The correlation of $V_{seebeck}$-$V_{heat}$ reflects the magnitude of $S$. This measurement was performed presently on the surface of the 0.8 μm thick SrNb$_{0.2}$Ti$_{0.8}$O$_3$/SrTiO$_3$ (001) with Ag$_2$Se and doped-CoSb$_3$ used as reference samples. Figure S6b shows the relationships of $V_{seebeck}$-$V_{heat}$ for the film and reference samples. It indicates a localised Seebeck coefficient (~290 μVK$^{-1}$) at the near-surface region (depth smaller than 100 nm) of the film material as shown in Figure 2c. This experiment demonstrates that the enhancement in $S$ was also observed at the surface of the film, which is in agreement with $S_{Film}$ as shown in Figure 2b.

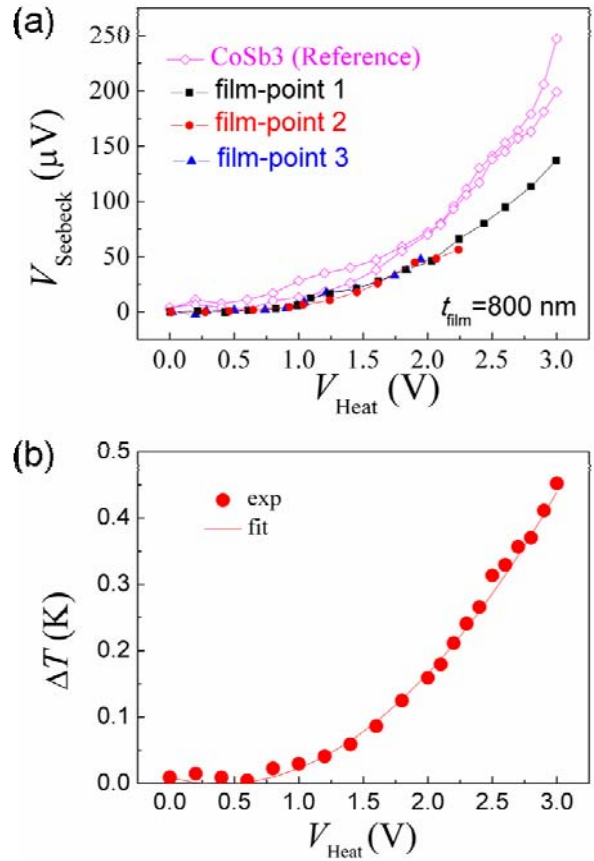

**Figure S6.** Localized characterizations of $S$ from the near-surface region (responding depth of ~100 nm) from a 0.8 μm thick film of $SrNb_{0.2}Ti_{0.8}O_3/SrTiO_3$ (001) by using nanometre-scale heating source in atomic force microscopy (AFM). **(a)** Relationship of the heating voltage ($V_{Heat}$) and the Seebeck voltage ($V_{Seebeck}$). **(b)** Relationship of the heating voltage ($V_{Heat}$) and the temperature difference ($\Delta T$).

**Section F: Converse confirmation for the present observed high TE performance**

Conversely, confirmation has been obtained by performing thermoshock (Figure S7) to eliminate the interfacial coherency and relax the interfacial strain for $SrNb_xTi_{1-x}O_3/SrTiO_3$ (001) with other compositions. As shown in Figure S7, performing five cycles of thermoshock upon the 130 nm thick $SrNb_{0.4}Ti_{0.6}O_3/SrTiO_3$ (001) at the deposition atmosphere significantly reduces both $S$ and $\sigma$. This result further demonstrates the understanding that the observed high TE performance is not dominated by interfacial diffusions or the conductance from reduced SrTiO$_3$ substrate. For example, the thermoshock performed at the deposition atmosphere is expected to promote the interfacial diffusion or generate more oxygen vacancy within the SrTiO$_3$ substrate.

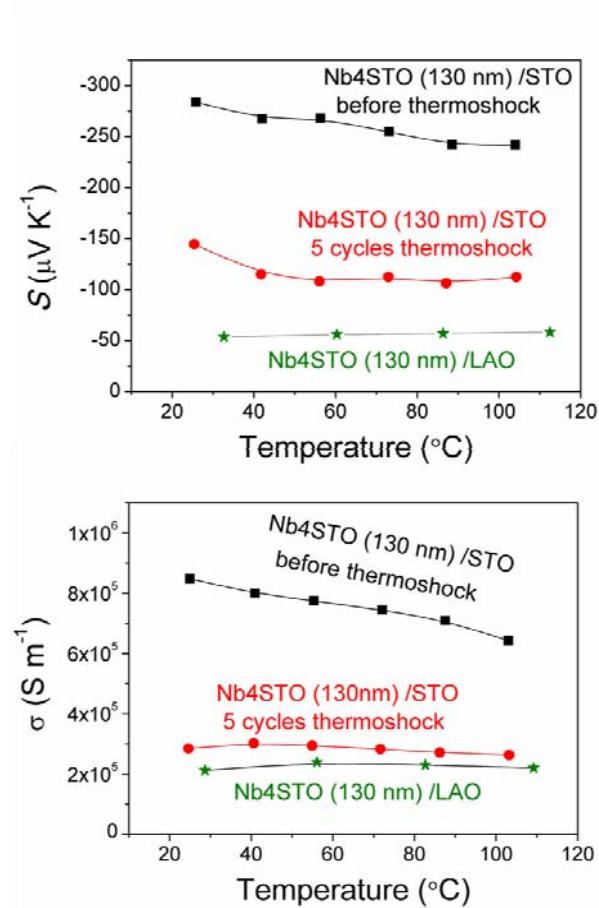

**Figure S7.** TE performance of 130 nm thick SrNb$_{0.4}$Ti$_{0.6}$O$_3$/SrTiO$_3$ (001) before and after five cycles of thermoshock and compared with the 130 nm thick SrNb$_{0.4}$Ti$_{0.6}$O$_3$/LaAlO$_3$ (001). The thermoshock was performed from 150 to 300 °C with a speed for temperature changing around 1 to 2 °C s$^{-1}$ at the film deposition atmosphere. The eliminations of the interfacial coherency and relax the interfacial strain are expected. Significant reductions in both $S$ and $\sigma$ are observed after the thermoshock and this conversely confirm the observed high TE

performance for $SrNb_xTi_{1-x}O_3$/$SrTiO_3$ (001).

We further examine the effect from compressive strain alone on the TE performance of $SrNb_xTi_{1-x}O_3$ by using coherently grown $SrNb_{0.2}Ti_{0.8}O_3$ and $SrNb_{0.4}Ti_{0.6}O_3$ on $DyScO_3$ (001) or $KTaO_3$ (001) single crystal substrate. The lattice constant of $DyScO_3$ ($a_0$ = 3.944 Å) and $KTaO_3$ ($a_0$ =3.989 Å) are in between of $SrTiO_3$ ($a_0$ =3.905 Å) and $SrNb_{0.4}Ti_{0.6}O_3$ ($a_0$ =3.989 Å). Therefore, a smaller lattice mismatch is expected when growing $SrNb_{0.2}Ti_{0.8}O_3$ and $SrNb_{0.4}Ti_{0.6}O_3$ on $DyScO_3$ (001) or $KTaO_3$ (001) substrates as compared with growing those on $SrTiO_3$ (001) substrates, which preserves the compressive distortion of film materials. This phenomenon has been further confirmed by XRD and RSM results for several samples as shown in Figure S8b to 8e. However, $S$ and $\sigma$ measured for compressively strained $SrNb_{0.4}Ti_{0.6}O_3$/$DyScO_3$ (001), $SrNb_{0.2}Ti_{0.8}O_3$/$DyScO_3$ (001) and $SrNb_{0.4}Ti_{0.6}O_3$/$KTaO_3$ (001) are similar to those from films grown on $LaAlO_3$ (001) substrate (strain relaxed) and the respective bulk values, as shown in Figure S8a. These results indicate that a compressive strain only cannot result in a similar high *TE* performance observed for $SrNb_xTi_{1-x}O_3$/ $SrTiO_3$ (001).

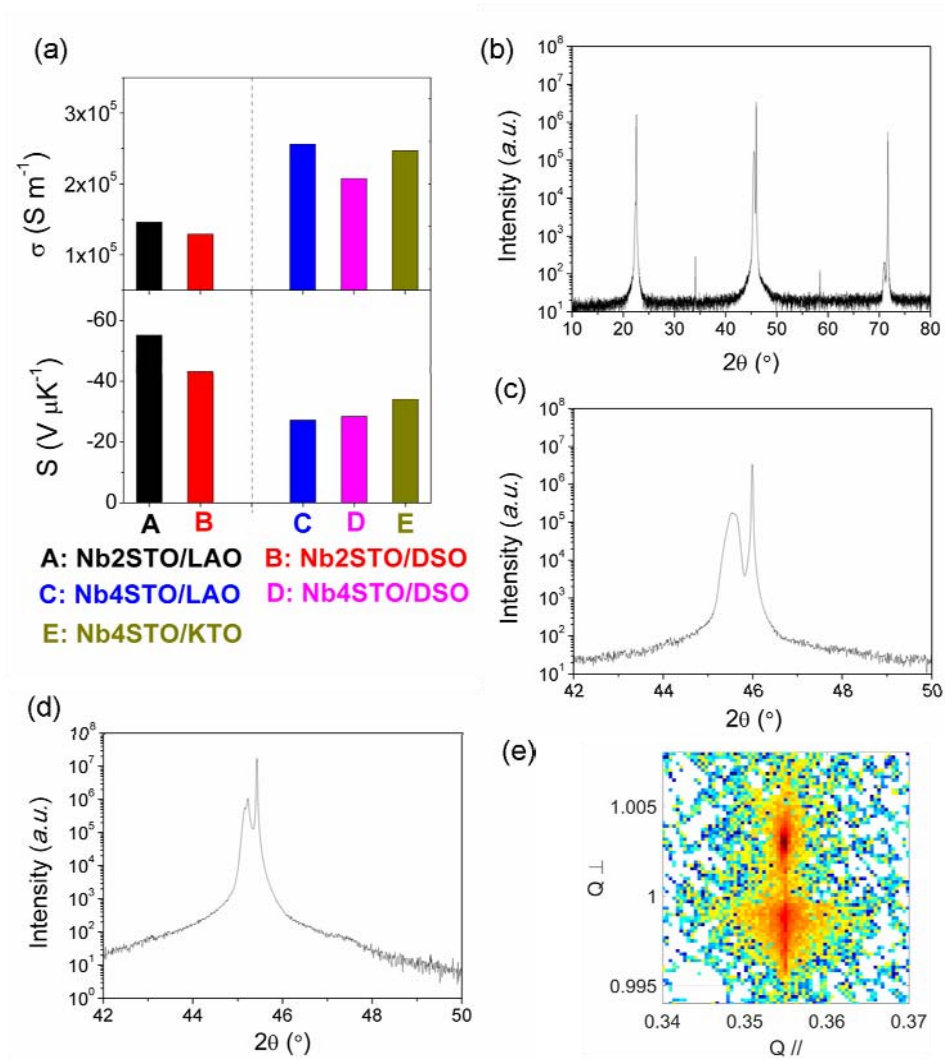

**Figure S8. (a)** $S$ and $\sigma$ measured for 400 nm thick **A**: $SrNb_{0.2}Ti_{0.8}O_3/LaAlO_3$ (001), **B**: $SrNb_{0.2}Ti_{0.8}O_3/DyScO_3$ (001), **C**: $SrNb_{0.4}Ti_{0.6}O_3/LaAlO_3$ (001), **D**: $SrNb_{0.2}Ti_{0.8}O_3/DyScO_3$ (001), E: $SrNb_{0.4}Ti_{0.6}O_3/KTaO_3$ (001), where B, D and E are compressively strained while A and C are strain-relaxed. No significant enhancements in $S$ and $\sigma$ are observed for B, D, E compared with A and C. The XRD diffraction patterns for $SrNb_{0.2}Ti_{0.8}O_3/DyScO_3$ (001) are shown in **(b)** and **(c)**. The XRD diffraction pattern for $SrNb_{0.4}Ti_{0.6}O_3/KTaO_3$ (001) is shown in **(d)**. The film and substrate possess the same crystal structure and orientation, since the diffraction peaks for the film are present beside those for the substrate **(b, c, d)**. The RSM result for $SrNb_{0.4}Ti_{0.6}O_3/KTaO_3$ (001) is shown in **(e)**, where the same in-plane projection in the reciprocal vector is observed between film and substrate.

**Section G: Further discussions on the possible mechanisms associated to the high TE performance**

Similar to AlGaN/GaN heterostructures[17,18,20], the ordering in polarons alignment is thermodynamically favoured by binding between interfacial and lattice polarization and is achieved under appropriate deposition kinetics. Figure S9a further illustrates a possible way to realize polarization induced electronic 2D confinement of carriers. Positive charge accumulated at the depletion region approaching to the film side, while the $Ti^+ \rightarrow O^-$ lattice dipoles generated in biaxial compressively distorted $SrNb_xTi_{1-x}O_3$ point towards the interface. Therefore, the ordering aligned lattice polarization and its induced periodical coulomb potentials along the cross-plane direction zigzag the conduction band across the Fermi energy and generate potential wells to realize electronic confinement of carriers as shown in Figure S9b.

The high relative permittivity of strain-distorted $SrTiO_3$ makes it practicable to realize above proposed electronic 2D confinement of carriers at the presently observed $n_{2D}$ by an aligned cross-plan lattice polarons. In Figure S9c and S9d, we further estimate the possibility to realize an electronic 2D confinement of carriers at different $n_{2D}$ by aligned cross-plan lattice polarons in strain-distorted $SrTiO_3$. This estimation is based on its reported dialectic properties reported by Ref. 23 as shown by Figure S11a. The capacity is written as: $C = \frac{\varepsilon_r \varepsilon_0 S}{d} = \frac{Q}{U}$, where $\varepsilon_0$ is the vacuum permittivity, $\varepsilon_r$ is the relative permittivity, $S$ is the planar area of the capacitor, $d$ is the interplaner distance of the capacitor, $Q$ is the accumulated charge and $U$ is the interplaner voltage. When the electrons from $n$ layers of unit cell of SrNb$_{0.2}$Ti$_{0.8}$O$_3$ are completely compressed to the surface layer, the accumulated 2D carrier concentration ($QS^{-1}$) is $n$ times 2.4x10$^{14}$ cm$^{-2}$ (the magnitude of presently $n_{2D}$ per in-plan unit layer) and $d$ equals to $n$ times 4 Å (the lattice constant). The required dialectic ability ($U$ x $\varepsilon_r$) is shown in Figure S10d. From Figure S10c, it can be seen that the allowed magnitude of $U$ x $\varepsilon_r$ for strain-distorted SrTiO$_3$ is around 10$^9$ to 10$^{10}$ Vm$^{-1}$. Such dialectic or ferroelectric performance can withstand complete separating carriers up to 10$^4$ layers of unit cells (several micrometre) as shown in Figure S10d without breaking down the material. This estimation indicates that an electronic 2D confinement of carriers at the presently observed $n_{2D}$ by an aligned cross-plan lattice polarons is practicable for a strain-distorted SrTiO$_3$ resulting from its high relative permittivity.

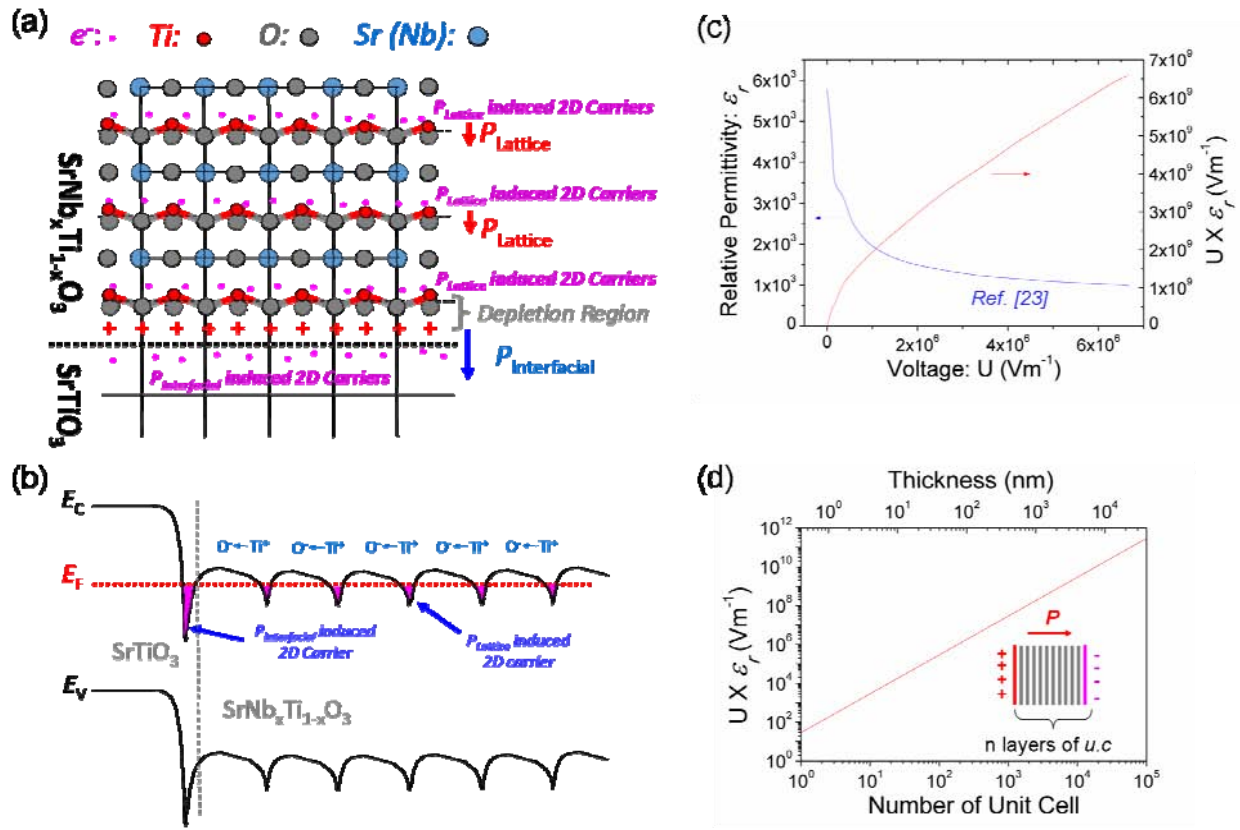

**Figure S9.** Schematic illustration of **(a)** the crystal structure and **(b)** the possible band structure to realize polarization induced electronic 2D confinement of carriers in $SrNb_xTi_{1-x}O_3/SrTiO_3$ (001). **(c)** Reported dialectic performance of strain-distorted $SrTiO_3$ at 10 GHz by ref. 23. **(d)** Estimation for required magnitude of voltage times relative permittivity ($U \times \varepsilon_r$) for completely compressing the electrons from $n$ layers of in-plane unit cell to the surface layer.

**References**

S1. Chen, J. Analysis of laser-induced plasmas utilizing $^{18}O_2$ as oxygen tracer. PhD Thesis, ETH Zurich (2014)

S2. Liu, H., Yuan, X., Lu, P., *et al.* Ultrahigh Thermoelectric Performance by Electron and Phonon Critical Scattering in $Cu_2Se_{1-x}I_x$. *Adv. Mater.* **25**, 6607-6612 (2013)